# Extracting Traceability Information from C# Projects

MICHAEL KERNAHAN, MIRIAM CAPRETZ, LUIZ CAPRETZ
Department of Electrical and Computer Engineering
University of Western Ontario
London, Ontario
CANADA
mkernaha@uwo.ca, mcapretz@eng.uwo.ca, lcapretz@eng.uwo.ca

*Abstract:* The maintenance portion of the software lifecycle represents a major drain on most software company's resources. The transition from programmers to maintainers is high risk, since usually the maintainers have to learn the system from scratch before they can begin modifying it appropriately. This paper introduces a method for automatically extracting important traceability information from a C# software project's source code. Using this traceability information, maintainers (and programmers) are better able to evaluate the impacts their actions will have on the entire project.

## 1 Introduction

Maintenance is an inescapable part of the software lifecycle process. Many organizations are realizing the importance of maintenance efforts to ensure the success of a software project. Maintenance efforts have become part of the overall software lifecycle process, with predetermined timelines and budgets. The acceptance of maintenance as an inevitable requirement has forced companies to examine their policies for hiring and training both their developers and their maintainers. Companies still have one major gripe with maintenance however: it means that they are losing money since the company rarely is able to generate income from maintenance efforts. For this reason, companies are interested in finding and exploiting ways that they can reduce maintenance effort and thus increase the profit margin on a software project.

One of the major hurdles of software maintenance is for the maintainers to actually understand the complete picture of what the original software does and how it does it. Without this thorough understanding, all but the simplest maintenance tasks can easily create more problems then they solve. Not only is it hard for the maintainer to determine if the fix has actually solved the problem, but it is almost impossible for the maintainer to ensure that no new bugs have been unsuspectingly added to the code. So the initial goal of maintenance efforts should be to read the technical documentation and try to understand the goals of the software and how all of the components fit together in the system. Unfortunately, as many maintainers can attest, technical documentation for software projects is usually lacking in depth and clarity.

Being able to visualize the traceability information for a software project in object-oriented programming develops with intimate knowledge of the system. Usually, the programmers who originally wrote the software understand what triggers events, and what handles those events. They also know the major variables within the major classes, and the implications of making changes to these variables. A maintainer or programmer new to the project will have a difficult time visualizing all of these relationships. This information can be very helpful in reducing improper usage of variables, and reducing duplicate effort.

This paper proposes a method for pulling out key information from C# source code directly. By removing the human factor from knowledge extraction from source code, it is hoped to produce useful information independently of coding style, comments, and any documentation that may have been produced. This method tokenizes the source code and creates an eXtensible Mark-up Language (XML) file that represents all of the *Namespaces*,

*Classes*, *Methods*, *Constructors*, *Variables*, *Events*, *Properties* and *Delegates* in the source file(s). This XML file is then parsed and utilised to create a database that can be used in a graphical user interface to view the traceability information for the code. In the remainder of this paper, Section 2 presents the problems that face maintenance efforts. Section 3 examines some of the research that is being done to address these issues. Section 4 reviews how traceability information is extracted from source code, while Section 5 presents the data model used to represent the information. Section 6 reviews the interface created for this work, while Section 7 suggests areas for future work in this field.

## 2 Problems for Maintenance

Technical documents written by programmers are usually too short and superficial, or too long and obtuse. On top of this, the documents are usually prepared once the software has been written already. This usually means that details are forgotten, and left out of the documentation. Most programmers on software projects have little or no experience with maintaining a software system that they did not help to create. Because of this, programmers rarely appreciate just how much of their knowledge about a software system is in their heads and not captured in documentation anywhere. This lack of understanding and the usual practice of creating documentation when a project is complete are major factors in the cost of maintenance.

Metrics can be useful for evaluating programs, as well as for trying to understand how a program was written. Managers can use metrics to track the performance of the development team, while programmers can use them to identify problem areas in the code. Metrics lose a lot of value when they are used without the context of how the project is put together and functions.

The use cases of software projects have a tendency to shift during the development stage. These shifts are often not recorded in the documentation, and usually undocumented use cases exist. Maintaining system integrity can be difficult without an understanding of the use cases of the system.

Throughout the development and maintenance stages, software projects are changed almost constantly. If such changes are implemented in an incomplete or inconsistent way, a loss of architectural quality will occur [1]. The lack of available traceability resources is a problem for collaboration during the development and maintenance phases of a software lifecycle.

## 3 Current Research

Vestdam and Nørmark [2,3], with their Elucidative documentation method, attempt to help programmers with documentation during the coding phase. Marks and Wilkie [4] present the OSCAR tool for extracting metrics from software automatically. Qin *et al.* [5] have studied extracting use cases from source code with this in mind.

Research is being done to look for ways to help the programmers maintain documentation throughout the software lifecycle. With ease of use, and more emphasis on the importance of documentation, solid documentation skills may develop in the industry. Unfortunately, there will always be a gap between what information the programmer records, and what the maintainers need.

Work in the field of traceability analysis for software projects has attempted to fill this gap with the information that the maintainers need. Riebisch [1] has begun work on a system to link design requirements to the actual source code. He has pointed out that many CASE tools could support traceability with minor amounts of effort.

Balzer and Deussen [6] have developed a graphical environment for representing the *package*, *class*, *method* and *attribute* levels of abstraction of Java code. Their Hierarchical Net is useful for seeing the tree like structure of a software system, and visually showing what method or class fits where.

DeLucia *et al.* [7] have acknowledged the tremendous time and effort required to produce meaningful traceability information manually or semi-automatically. They have proposed a solution for finding traceability links between software artefacts. In their solution, both the software engineer and the system identify links. These two groups of links are analysed to find *candidate links* and *warning links*, which may need to be added or removed from the system.

## 4 Traceability Extraction

The first task to address when automating knowledge extraction from source code is to understand how the code will be interpreted into tokens. To accomplish this task, a tokenizer developed by the #Develop[1] open source project has been used. This initial tokenization was performed without making any changes to the tokeniser developed for the #Develop communities' tool which is able to convert C# code to VB.Net code [8] and vice versa.

Once the source code had been parsed and tokenized, the next step was to represent this code in an XML format so that it can be used by other tools, as well this one. Unfortunately, because of the complexity of the code, it was not possible to simply use the serialisation capabilities of object-oriented programs. Serialisation is essentially the automatic mapping of objects into binary or XML files. Since serialisation was not possible with the tokenised data model, the #Develop C# to VB converter was used as a starting point for creating the XML output. Instead of outputting clean VB code, the methods were rewritten to generate XML nodes. Currently this step strips out much of the information from the source code. Since the focus of this tool is not to evaluate metrics, but rather to extract traceability information, statements such as *if*, *else*, *for*, *while*, *switch*, *case*, etc. are not relevant. The variables used in these statements are recorded, but the overall structure within methods was not. Future work on this project may look at extracting metrics to evaluate not only the traceability information, but also to provide a report on the quality of the code.

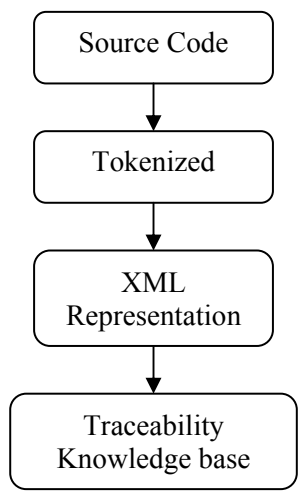

Fig. 1: Information flow

With the code transformed into an XML representation, all that remains is to pull out the information that we are interested in, and then represent it in such a way that the traceability information can be visually understood. A data model was created to represent the traceability information. The XML code representation was then parsed in order to extract the information and populate the data model. Figure 1 shows the flow of information going from source code to the traceability knowledge base.

## 5 Data Model
The data model created for this project needed to represent the traceability information for numerous different types of objects, with numerous different types of relationships. It was decided to design the data model in such a way that new categories of object types or relationships could be added to the knowledge base at any time. At the highest level, there are five major object types in the data model. The relationships between the top-level object types are shown in Figure 2. These types and their relationships are explained below.

### 5.1 KnowledgeBase Object
The topmost element in the data model, the KnowledgeBase object is essentially a wrapper that contains two lists of sub-objects. These lists contain KnowledgeType and LinkType objects, both of which are explained below. All changes made to the data model are sent to the KnowledgeBase object, which then sends out events to the listeners (the user interface mostly).

### 5.2 KnowledgeType Object
Every object that we are trying to represent in our traceability system needs to have a specific type. The current list of types developed for this system is: *Namespace*, *Class*, *Constructor*, *Method*, *Property*, *Variable*, *Delegate*, and *Event*. A KnowledgeType object represents each of these types. KnowledgeType objects point directly to all of the top level KnowledgeObjects that they contain. The user interface represents each KnowledgeType object as a column in the interface. These columns may be displayed or hidden, and can be displayed in any order. Queries to the knowledge base can be based on the KnowledgeType object.

---
[1] #Develop – www.icsharpcode.net

## 5.3 KnowledgeObject Object
The KnowledgeObject object is used to represent each object that is extracted from the source code. Every KnowledgeObject must have a specific KnowledgeType, and thus will be represented in that column in the interface. Each KnowledgeObject also contains a list of LinkObjects, which are used to represent the relationships between the KnowledgeObjects. After the automated extraction of traceability information has taken place, the user is able to add extra information to KnowledgeObjects. This information may include adding notes about important information or problems associated with the KnowledgeObject, or links to documents that describe the object.

## 5.4 LinkObject Object
All of the relationships among the different KnowledgeObject objects are represented as LinkObject objects. Each LinkObject object has a parent and child KnowledgeObject object, used to represent the directionality of the relationship. The LinkObject also specifies the LinkType of the relationship, which essentially describes the relationship between the two KnowledgeObject objects.

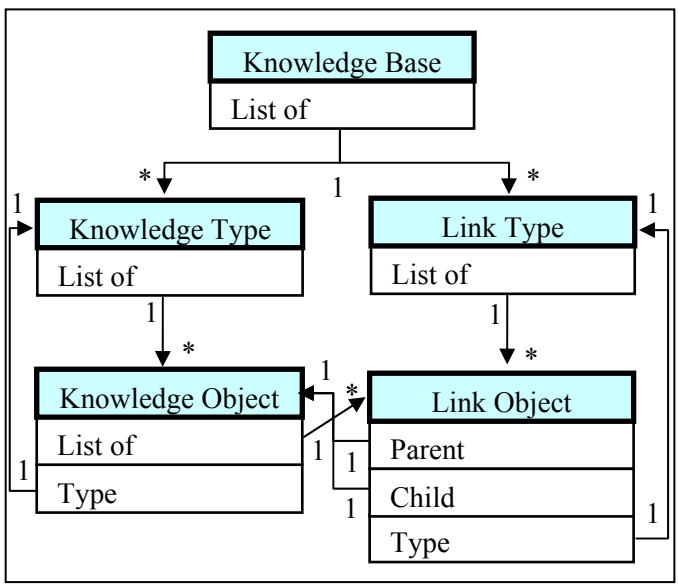

Fig. 2:  Top level objects and their relationships

## 5.5 LinkType Object
As previously mentioned, every relationship in the traceability system involves two KnowledgeObject objects, which are linked together by a specific LinkObject object. Every LinkObject object has a specific LinkType object, which essentially describes the relationship between the two KnowledgeObject objects. By having these separate LinkObject and LinkType objects, the relationships between the different KnowledgeObject objects can be distinguished from each other. This allows the user to display traceability information for only certain LinkTypes if they chose.

## 6 User Interface
The main goal of the user interface is to represent the traceability information in a way that is easily understandable. The success of many software projects depends on the user interface. In order to clarify the different object types that are represented by the system, each object type is represented by its own column in the interface. Columns may be added or removed from the interface, as well as have their order changed. Adding a new column, for example a *Requirements* column, is as simple as creating a new KnowledgeType object in the data model. Part of the user interface is shown in Figure 3.

Every node in the different columns has a checkbox in front of it. Checking off this check box will cause the system to access the data model and determine the traceability information related to the (un)selected node. This traceability information is represented visually onscreen by parsing the lists of objects in the other columns. If a node has been checked off, then all of the objects that are related to that node are displayed in their respective column, while nodes that are not related to any of the selected nodes will not be shown. This traceability information is passed from column to column, so objects that are indirectly related to the selected object will also be displayed. If a column has no nodes checked off, then the traceability for all of the displayed nodes will be used, but once a node has been checked off, only the traceability information for the checked off nodes will be used.

The traceability information is also represented in tree format within the columns themselves. The user can jump to any item of interest in the software project, and expand the node. This will reveal all of the traceability information where the expanded node is the parent of the relationship. Figure 4 shows a small example of what the traceability tree could look like. In Figure 3, the Class, Method and Variable sections are shown at the top level. The *SplitVertex* node, which has been expanded, shows how the traceability information is shown in both tree and column format for easy accessibility.

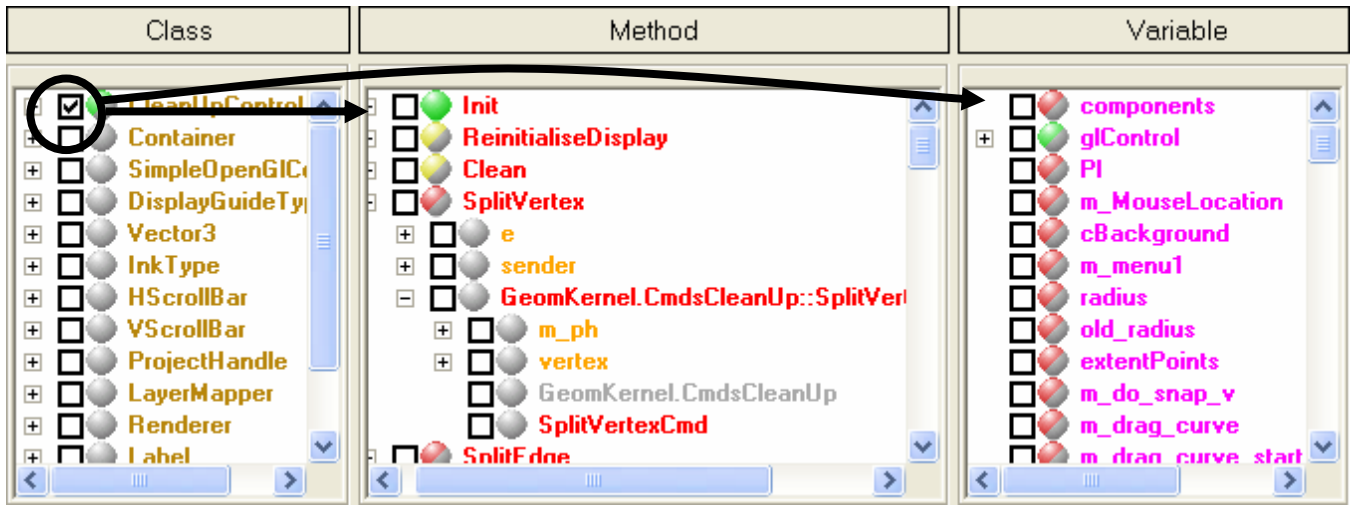

Fig. 3: User Interface

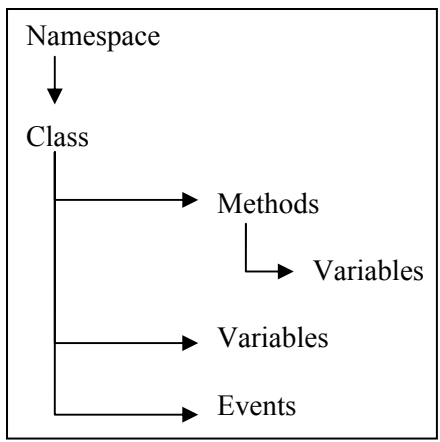

Fig. 4: Traceability Tree

Different colours of text are used to represent different types of objects. In the example in Figure 3, red is used to denote *Methods* and *Method Calls* (such as *Init* and *SplitVertex*), while orange is used for *Parameters* (such as *e* and *sender*) that are passed to the *Methods*. *Variables* (such as *glControl* and *m_drag_curve*) are magenta, and *Namespaces* (such as *GeomKernel.CmdsCleanUp*) are grey. Types not represented in Figure 3 also have distinct colours. The coloured balls are used to represent the accessibility level of the node and its children. Green is used for public, red is used for private, and yellow is used for all other levels.

Once the automatic extraction has populated the knowledge base with traceability information, the user is free to modify the resulting data model any way that they choose. Simply simply dragging one object onto another object will create a relationship between them. At the bottom of the interface, a panel exists for adding extra information about the selected object.

## 7 Future Work

This prototype has yet to be used in a case study. The first obvious future step is to have a development team use the tool on a real world project. From the team's input, it can be determined if the tool was useful, and the team's comments and suggestions can help direct the future work of this project.

As mentioned previously, future work on this project may include adding metric extraction functionality. The metrics could be represented as attributes, and visual representation could also be achieved using different colours or graphics to represent the nodes. Microsoft's Visual Studio 2005 may include the use of OOML, a mark-up schema for object oriented programming languages. If this occurs, then this tool would be able to work from the source code in real time, and perhaps be integrated into the Visual Studio IDE. Traceability information, combined with rules about objects in the source code, could flag problems for the developer before they even try to compile the code. Experience shows that real time functionality is pretty much required to gain acceptance from the development community. It is for this reason that this will be the major focus of future work.

The current major hurdle with the prototype is the level of interconnectedness between the different nodes. This interconnectedness can reduce the usability of the traceability information since for some projects virtually every method can be linked to any other method through other intermediate methods. Other points of interest include pulling out the comments from source code and attaching them to the traceability nodes, as well as finding a way to capture requirements for the project and mapping them to traceability nodes.

# 8 Conclusion

This paper has described a traceability tool developed for C# software projects. The traceability information extracted from project source code can be very useful for team members (programmers or maintainers) who are new to a software project. Generally, the documentation required to fully understand a software project is not available until the development has been completed, and even then it is often not complete. By being able to visually see the traceability information, developers are able to quickly evaluate the impact that changes they introduce will have on the entire system, as well as being able to track bugs and finding where variables are being updated by the system. Using tools such as this one and the others discussed in Section 3, the future work of maintainers should become easier, faster, and involve much less risk in a business sense. The transition times during which maintainers must learn how the software system works will continue to fall as tools such as this one are integrated into real time IDE interfaces used for development.

*References:*
[1]  M. Riebisch, Supporting Evolutionary Development by Feature Models and Traceability Links, *Proceedings of the 11th International Conference and Workshop on the Engineering of Computer-Based Systems, ECBS 2004*, May 24–27, 2004, pp. 370–377.

[2]  T. Vestdam, K. Nørmark, Aspects of Internal Program Documentation – an Elucidative Perspective, *Proceedings of the 10th International Workshop on Program Comprehension*, June 26–29, 2002, pp. 43–52.

[3]  T. Vestdam, Elucidative Programming in Open Integrated Development Environments for Java, *Proceedings of the 2nd International Conference on Principles and Practice of Programming in Java*, June 16–18, 2003, pp. 49–54.

[4]  R. M. Marks, F. G. Wilkie, Visualising Object-Oriented Source Code Complexity using XML, *Proceedings of the 9th IEEE International Conference on Engineering Complex Computer Systems Navigating Complexity in the e-Engineering Age*, April 14–16, 2004, pp. 161–170.

[5]  T. Qin, L. Zhang, Z. Zhou, D. Hao, J. Sun, Discovering Use Cases from Source Code using the Branch-Reserving Call Graph, *Proceedings of the 10th Asia-Pacific Software Engineering Conference*, December 10–12, 2003, pp. 60–67.

[6]  M. Balzer, O. Deussen, Hierarchy based 3D Visualization of Large Software Structures, *Proceedings of the IEEE Conference on Visualization, VIS'04*, October 10–15, 2004, p. 4.

[7]  A. DeLucia, F. Fasano, R. Oliveto, G. Tortora, Enhancing an Artefact Management System with Traceability Recovery Features, *Proceedings of the 20th IEEE International Conference on Software Maintenance, ICSM 2004*, September 11–17, 2004, pp. 306–315.

[8]  #Develop Open Source Community, "C# to VB.NET Converter", http://www.icsharpcode.net/OpenSource/SD/Download/